\documentclass[12pt]{iopart}

\usepackage{graphicx}
\usepackage{tipa}
\usepackage{amssymb}
\usepackage{txfonts}
\usepackage{hyperref}
\usepackage{graphicx}
\usepackage{epsfig}
\usepackage{bm}
\begin{document}
\title{Massive scalar field quasinormal modes of a Schwarzschild black hole
            surrounded by quintessence}

\author{Chunrui Ma and Yuanxing Gui, Wei Wang, Fujun Wang}

\address{School of Physics and Optoelectronic Technology,\\ Dalian, 116024, People's Republic of China\\machunrui0312@163.com and thphys@dlut.edu.cn}

\begin{abstract}
We present the quasinormal frequencies of the massive scalar field
in the background of a Schwarzchild black hole surrounded by
quintessence with the third-order WKB method. The mass of the
    scalar field $u$ plays an important role in studying the quasinormal
    frequencies, the real part of the frequencies increases linearly as mass $u$
    increases, while the imaginary part in absolute value decreases linearly which leads to damping more slowly and the frequencies having a limited value. Moreover, owing to the
presence of the quintessence, the massive scalar field damps more
slowly.

\end{abstract}

\maketitle

\section{Introduction}
The quasinormal modes of black holes have drawn much attention in
recent years. Vishveshwara first put forward the concept of
quasinormal modes in calculations of the scattering of
gravitational radiation by a Schwarzschild black-hole\cite{1} and
Press proposed the term quasinormal frequencies\cite{2}. The
quasinormal frequencies are an important characteristic of a black
hole, because the frequencies only depend on the black hole
parameters rather than the initial perturbation. In addition, the
properties of quasinormal modes have been explored in the Ads/CFT
correspondence\cite{3} and loop quantum gravity\cite{4}.

On the other hand, a large number of astronomical observations,
such as type Ia supernovae \cite{5}, CMB\cite{6} and large scale
structure\cite{7},
 indicate that the expansion of the universe is speeding up rather than
 slowing down. To explain this accelerated expansion, the
 universe is regarded as being dominated by an exotic component
    with large negative pressure called "dark
 energy" which
constitutes about 70\% of the energy density of the universe.
There are several candidates for dark energy: the cosmological
constant\cite{8} and dynamic candidates(for example
phantom\cite{9}, quintessence\cite{10}, k-essence\cite{11} and
quintom\cite{12}). The difference of these candidates for dark
energy is the size of the parameter $\omega_{q}$, namely, the
radio of the pressure and energy density of the dark energy and
for quintessence
 $-1\leq\omega_{q}\leq\frac{-1}{3}$.

The quasinormal modes of different fields perturbation around
different black holes have been widely
investigated\cite{13}-\cite{32}, especially the massless scalar
field\cite{33}. Although the massive field was studied in
different black holes\cite{34}-\cite{37}, the massive field
quasinormal modes still have gaps. Kiselev\cite{38} recently
considered Einstein's field equations for a black hole surrounded
by quintessence and obtained a new solution related to state
parameter $\omega_{q}$ of the quintessence. In this paper, we use
the third-order WKB method to explore the quasinormal modes of
massive scalar field perturbation around a Schwarzschild black
hole surrounded by quintessence.

The outline of this paper is as follows: in section 2, we evaluate
 the quasinormal
frequencies of the low overtones quasinormal modes. The discussion
and summary are presented in section 3.

\section{Massive scalar field quasinormal modes  of a Schwarzschild black hole surrounded by Quintessence}
Kiselev\cite{38} stated a new static spherically-symmetric exact
solution of Einstein equations describing a black hole charged or
not and surrounded by the quintessence under the condition of
additivity and linearity in the energy-momentum tensor. For the
Schwarzschild black hole, the metric is given by\cite{39}:
\begin{equation}
ds^{2}=(1-\frac{2M}{r}-\frac{c}{r^{3\omega_{q}+1}})dt^{2}-(1-\frac{2M}{r}-\frac{c}{r^{3\omega_{q}+1}})^{-1}dr^{2}-r^{2}(d\theta^{2}+\sin\theta^{2}d\phi^{2})
,\,
\end{equation}
where $M$ is the mass of the black hole, $\omega_{q}$ is the
quintessential state parameter, $c$ is the normalization factor
dependent on
$\rho_{q}=-\frac{c}{2}\frac{3\omega_{q}}{r^{3(1+\omega_{q})}}$,
and $\rho_{q}$ is the density of quintenssence.

The massive scalar field in a curved background is governed by the
Klein-Gordon equation:
\begin{equation}
\Box{\Phi}-u^{2}\Phi=\frac{1}{\sqrt{-g}}(g^{uv}\sqrt{-g}\Phi_{,u})_{,v}-u^{2}\Phi=0,
\end{equation}
where $\Phi$ is the scalar field.

After substituting equation(1) into equation(2) and separating
angular and time variables, we obtain the radial equation:
\begin{equation}
(\frac{d^{2}}{dr_{*}^{2}}+\omega^{2}-V(r))\Phi(r)=0,
\end{equation}
where:
\begin{equation}
V(r)=(1-\frac{2M}{r}-\frac{c}{r^{3\omega_{q}+1}})(\frac{l(l+1)}{r^{2}}+\frac{2M}{r^{3}}+\frac{c(3\omega_{q}+1)}{r^{3\omega_{q}+3}}+u^{2}),
\end{equation}
\begin{equation}
dr_{*}=\frac{dr}{1-\frac{2M}{r}-\frac{c}{r^{3\omega_{q}+1}}},
\end{equation}
and $l=0,1,2,3...$ parameterizes the field angular harmonic index.
The effective potential $V(r)$ approaches to a constant both at
the event horizon and at spatial infinity. It is clear that the
effective potential relates to the value of $r$, angular harmonic
index $l$, the state parameter $\omega_{q}$, the scalar field mass
$u$, the normalization factor $c$ and the mass of the black hole
$M$. However, in this paper, we only want to investigate the
relationship between the state parameter $\omega_{q}$ or the
scalar field mass $u$ and the quasinormal modes. Therefore, taking
$M=1$ and $c=0.001$, we compute the quasinormal frequencies
stipulated by above potential using the third-order WKB method
developed by Schutz, Will and Iyer\cite{40}-\cite{42}:
\begin{equation}
\omega^{2}=[V_{0}+(-2V_{0}^{''})^{1/2}\Lambda]-$i$(n+\frac{1}{2})(-2V_{0}^{''})^{1/2}(1+\Omega),
\end{equation}
where
\begin{eqnarray}
\Lambda=\frac{1}{(-2V_{0}^{''})^{1/2}}\{\frac{1}{8}(\frac{V_{0}^{(4)}}{V_{0}^{''}})(\frac{1}{4}+\alpha^{2})-\frac{1}{288}(\frac{V_{0}^{'''}}{V_{0}^{''}})^{2}(7+60\alpha^{2})\},\\
\Omega=\frac{1}{(-2V_{0}^{''})}\{\frac{5}{6912}(\frac{V_{0}^{'''}}{V_{0}^{''}})^{4}(77+188\alpha^{2})\nonumber\\-\frac{1}{384}(\frac{V_{0}^{'''^{2}}V_{0}^{(4)}}{V_{0}^{''^{3}}})(51+100\alpha^{2})
+\frac{1}{2304}(\frac{V_{0}^{(4)}}{V_{0}^{''}})^{2}(67+68\alpha^{2})\nonumber\\+\frac{1}{288}(\frac{V_{0}^{'''}V_{0}^{(5)}}{V_{0}^{''^{2}}})(19+28\alpha^{2})-\frac{1}{288}(\frac{V_{0}^{(6)}}{V_{0}^{''}})(5+4\alpha^{2})\},
\end{eqnarray}
and
\begin{equation}
\alpha=n+\frac{1}{2},
\end{equation}
\begin{equation}
 V_{0}^{(n)}=\frac{d^{n}V}{dr_{*}^{n}}|_{r_{*}=r_{*}(r_{p})},
\end{equation}
where $n$ is the overtone number

Substituting the effective potential (4) into the formula above,
we can get the quasinormal frequencies for the massive scalar
field in the Schwarzschild black hole surrounded by quintessence
background and the quasinormal frequencies are shown in table 1,
table 2, table3 and figures 1-2.

\begin{table}
\caption{\label{arttype}Values of the quasinormal frequencies for
low overtones in the Schwarzschild black hole without
quintessence($c=0$) for fixed $l=5,u=0.2$ } \footnotesize\rm
\begin{tabular*}{\textwidth}{@{}l*{15}{@{\extracolsep{0pt plus12pt}}l}}
\br
$\omega(n=0)$&$\omega(n=1)$&$\omega(n=2)$&$\omega(n=3)$&$\omega(n=4)$\\
\mr \ 1.065754-0.095396i &1.055414-0.287524i
 &1.036163-0.483127i&1.010110-0.683090i&0.9790981-0.887095i\\

\br
\end{tabular*}
\end{table}

\begin{table}
\caption{\label{arttype}Values of the quasinormal frequencies  for
low overtones in the Schwarzschild black hole surrounded by
quintessence($c=0.001$) for fixed $l=5,u=0.2$ } \footnotesize\rm
\begin{tabular*}{\textwidth}{@{}l*{15}{@{\extracolsep{4pt plus12pt}}l}}
\br
$3\omega_{q}+1$&$\omega(n=0)$&$\omega(n=1)$&$\omega(n=2)$&$\omega(n=3)$&$\omega(n=4)$\\
\mr
\ 0.0 &1.064163-0.095203i & 1.053847-0.286942i &1.034639-0.482147i &1.008645-0.681703i&0.977700-0.885292i\\
-0.2&1.063768-0.095151i&1.053458-0.286786i&1.034262-0.481884i&1.008282-0.681329i&0.977355-0.884806i\\
-0.4&1.063274-0.095089i&1.052972-0.286597i&1.033790-0.481566i&1.007828-0.680878i&0.976921-0.884220i\\
-0.6&1.062656-0.095014i&1.052364-0.286373i&1.033200-0.481186i &1.007261-0.680339i &0.976380-0.883517i\\
-0.8&1.061885-0.094927i &1.051606-0.286107i &1.032465-0.480738i &1.006556-0.679700i &0.975710-0.882683i\\
-1.0&1.060920-0.094825i &1.050659-0.285798i &1.031552-0.480214i &1.005685-0.678952i &0.974885-0.881701i\\
-1.2&1.059714-0.094710i &1.049480-0.285446i &1.030420-0.479613i &1.004615-0.678084i &0.973881-0.880551i\\
-1.4&1.058205-0.094581i &1.048013-0.285053i &1.029026-0.478931i &1.0033103-0.67709i &0.972671-0.879211i\\
-1.6&1.056320-0.094443i &1.046191-0.284623i &1.027316-0.478172i &1.001734-0.675948i &0.971230-0.877652i\\
-1.8&1.053963-0.094300i &1.043933-0.284170i &1.025229-0.477343i &0.999850-0.674662i &0.969539-0.875834i \\
-2.0&1.051018-0.094162i &1.041143-0.283715i&1.022701-0.476462i &0.997622-0.673216i &0.967584-0.873701i\\
\br
\end{tabular*}
\end{table}

The date of table 1 is the quasinormal frequencies of a
Schwarzschild black hole without quintessence and under the
quintessence is given in table 2. Figure 1 shows that the real
part and the imaginary part of the quasinormal frequencies  change
as the quintessential state parameter $\omega_{q}$ changes for
fixed mass $u$. Comparing the table 1 with the table 2, we find
the real part and the magnitude of imaginary part in the
Schwarzschild space-time without quintessence are larger. It means
that due to the presence of quintessence, the oscillations damp
more slowly. Furthermore, the imaginary part in absolute value and
the real part decrease as the value of $\omega_{q}$ decreases, as
shown in figure 1 and table 1.

\begin{figure}[th]
\centerline{\psfig{file=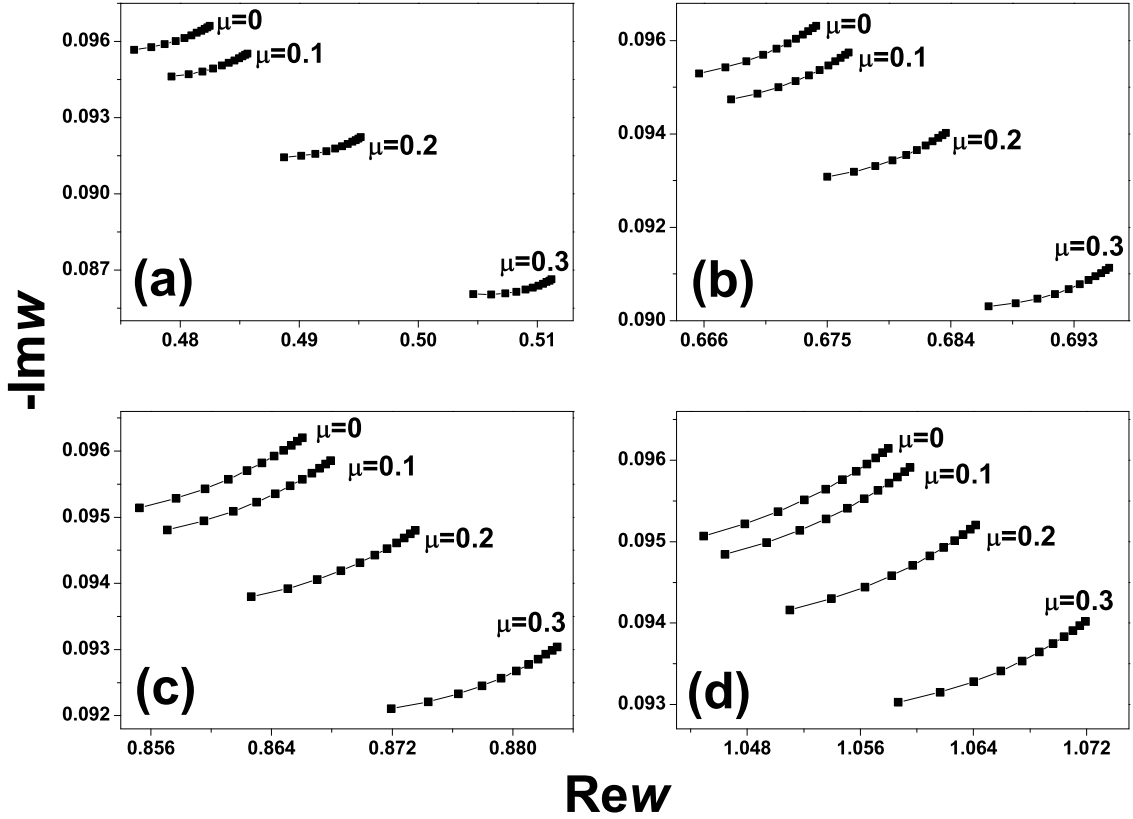,width=6.8in}} \vspace*{8pt}
\caption{quasinormal frequencies of the black hole surrounded by
quintessence for $c=0.001, n=0, u=0,0.1,0.2,0.3, 3\omega_{q}+1$
runs the values from 0 to  -2.0 at intervals of -0.2.  (a):
$l=2$,(b):$l=3$, (c):$l=4$,(d):$l=5$. } \label{fig1}
\end{figure}

We present the quasinormal frequencies for different values of the
mass of scalar field $u$ in figs 2 and table 3. Notice that, The
real part of the quasinormal frequencies grows with increasing of
the mass field $u$, while the imaginary part of quasinormal
frequencies in absolute value falls down. Moreover, the
frequencies change linearly as $u$ changes as shown in figure 2.

\begin{table}
\caption{\label{arttype}Values of the quasinormal frequencies for
fixed $l=3$,$n=0$. in the Schwarzschild black hole surrounded by
quintessence($c=0.001$) for different values of mass $u$.}
\footnotesize\rm
\begin{tabular*}{\textwidth}{@{}l*{15}{@{\extracolsep{4pt plus12pt}}l}}
\br
$3\omega_{q}+1$&$u=0$&$u=0.1$&$u=0.2$&$u=0.3$\\
\mr
\ 0.0&0.674185-0.096319i &0.676547-0.095748i &0.683650-0.094028i &0.695548-0.091132i \\
-0.2&0.673930-0.096267i &0.676292-0.095696i &0.683395-0.093977i & 0.695291-0.091082i\\
-0.4&0.673611-0.096204i &0.675973-0.095634i & 0.683074-0.093915i &0.694968-0.091022i\\
-0.6&0.673212-0.096129i &0.675572-0.095559i &0.682671-0.093842i &0.694561-0.090952i\\
-0.8&0.672711-0.096041i &0.675071-0.095472i&0.682167-0.093757i&0.694051-0.090871i\\
-1.0& 0.672084-0.095940i &0.674442-0.095372i &0.681534-0.093660i &0.693411-0.090780i\\
-1.2&0.671299-0.095825i &0.673655-0.095258i &0.680740-0.093552i & 0.692607-0.090680i\\
-1.4&0.670317-0.095698i &0.672670-0.095134i &0.679746-0.093434i &0.691597-0.090575i\\
-1.6&0.669089-0.095562i &0.671438-0.095001i &0.678501-0.093312i &0.690331-0.090471i\\
-1.8&0.667553-0.095425i &0.669897-0.094868i &0.676944-0.093193i &0.688745-0.090377i\\
-2.0&0.665636-0.095295i &0.667972-0.094745i &0.674997-0.093089i &0.686761-0.090308i\\

\br
\end{tabular*}
\end{table}

\begin{figure}[th]
\centerline{\psfig{file=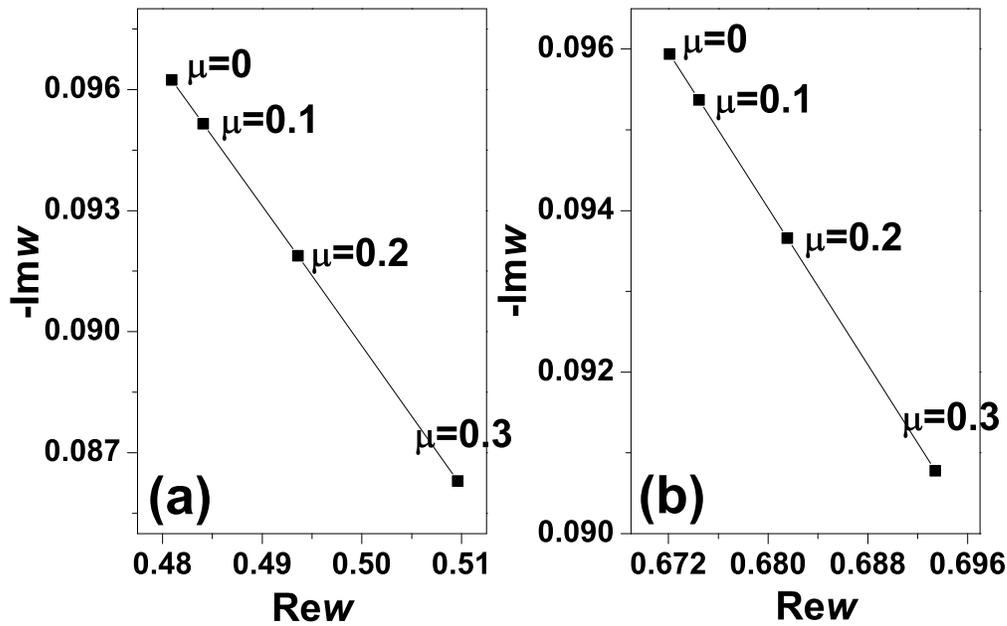,width=6.0in}} \vspace*{8pt}
\caption{quasinormal frequencies  of the black hole surrounded by
quintessence for $c=0.001, n=0, u=0,0.1,0.2,0.3,
3\omega_{q}+1=-1.0$  (a): $l=2$, (b):$l=3$  } \label{fig1}
\end{figure}

\section{Discussion and summary}
We have thoroughly investigated the quasinormal modes for massive
scalar field perturbation in a Schwarzschild black hole surrounded
by quintessence background. The paper proposes the quasinormal
modes are greatly influenced by the quintessence and the mass of
scalar field, because the introduction of the quintessence and the
mass $u$ leads to less damping of the quasinormal modes. Actually
$c$ may be too smaller than 0.001 to neglect the influence of the
quinntessence. However, if the density of quintessence surrounding
the black hole is high enough to influence distinctly the
quasinormal modes, we can study the character of quintessence by
the experimental date.

Another new phenomena found here is for given $l$, $n$, and
$\omega_{q}$, the real part of the frequencies linearly increase,
while the magnitude of imaginary part linearly decrease as the
mass of the scalar field $u$ increases. As we know, the mass of
the scalar field $u$ has a maximum value dependent on the mode
under consideration\cite{43}. That is mean, the quasinormal
frequencies have a limited value. Thereby, the introduction of the
the quintessence and the mass $u$ has enriched the quasinormal
frequencies of the Schwarzschild black hole.

\section{Acknowledgments}
I would like to acknowledge Zhanhui Wang and Ming Liu for helpful
discussions. This work is supported by the National Natural
Science Foundation of China under Grant No. 10573004.


\section*{References}
\begin{thebibliography}{10}
\bibitem{1} Vishveshwara C V 1970 \emph{Nature} \textbf{227} 936
\bibitem{2} Press W H 1971 \emph{Astrophys.J.} \textbf{170} L105
\bibitem{3} Maldacena J M 1998 \emph{Adv.Theor.Math.Phys.}
            \textbf{2} 231 \\Gubser S S 2001 \emph{Phys.Rev.D} \textbf{63}
              084017 \\Cardoso V and Lemos J P S 2001 \emph{Class.Quantum.Grav.}
              \textbf{18} 5257\\ Birmingham D, Sachs I and Solodukhin S N 2002
              \emph{Phys.Rev.Let.} \textbf{88} 151301\\Konoplya R A 2002
             \emph{Phys.Rev.D} \textbf{66} 044009\\Wang B \emph{et al} 2004
                \emph{Phys.Rev.D} \textbf{70} 064025
\bibitem{4} Hod S 1998 \emph{Phys.Rev.Let.} \textbf{81}
4293\\Corichi A 2003 \emph{Phys.Rev.D} \textbf{67} 087502\\Dreyer
O 2003  \emph{Phys.Rev.Let.} \textbf{90} 081301
\bibitem{5}Perlmutter S et al 1999 \emph{Astrophys.J.}\textbf{517}
        565\\
        Knop R A et al 2003 \emph{Astrophys.J.}\textbf{598} 102\\
        Riess A G et al 2004 \emph{Astrophys.J.}\textbf{607} 665\\
        Zhang X and Wu F Q 2005 \emph{Phys.Rev.D} \textbf{72}
        043524
\bibitem{6}  De Bernardis P \emph{et al} 2000 \emph{Nature}
        \textbf{404} 955\\ Halverson N W \emph{et al} 2002
        \emph{Astrophys.J.} \textbf{568} 38\\ Lamon R
        and Durrer R 2006 \emph{Phys.Rev.D} \textbf{73} 023507
\bibitem{7} Bacon D J \emph{et al} 2000 \emph{Mon.Not.R.Astron.Soc.} \textbf{318} 625
            \\ Bacon D J \emph{etal} 2003 \emph{Mon.Not.R.Astron.Soc.} \textbf{344} 673
            \\Tegmark M\emph{et al} 2004 \emph{Phys.Rev.D} \textbf{69} 103501
\bibitem{8} Padmanabhan T 2003 \emph{Phys Rep} \textbf{380} 235
             \\Alcaniz J S 2004 \emph{Phys.Rev.D} \textbf{69} 083521
\bibitem{9}Caldwell R R 2002 \emph{Phys.Lett.B} \emph{545} 23
            \\Chimento L P and Lazkoz R 2003 \emph{Phys.Rev.Lett.}
            \textbf{91} 211301\\ Vikman A 2005 \emph{Phys.Rev.D}
            \textbf{71} 023515
\bibitem{10} Caldwell R R \emph{et al}
             1998 \emph{Phys.Rev.Lett.} \textbf{80} 1582\\ Sahni V and
            Wang L M 2000 \emph{Phys.Rev.D} \textbf{62} 103517 \\
            Capozziello S \emph{et al} 2006 \emph{Class.Quantum.Grav.} \textbf{23}
            1205\\ Rogerio Rosenfeld and Joshua A Frieman 2006
            astro-ph/0611241V1
\bibitem{11} Chiba T \emph{et al} 2000 \emph{Phys.Rev.D}
             \textbf{62} 023511\\ Scherrer R J 2004 \emph{Phys.Rev.Lett.}
             \textbf{93} 011301
\bibitem{12} Wei H \emph{et al} 2005 \emph{Class.Quantum.Grav.}
             \textbf{22} 3189 \\ Zhao G B \emph{et al} 2005 \emph{Phys.Rev.D}
            \textbf{72} 123515 \\ Feng B \emph{et al} 2006
            \emph{Phys.Lett.A} \textbf{20} 2075

\bibitem{13} Regge T and Wheeler J A  1957 \emph{Phys.Rev.} \textbf{108}1063
\bibitem{14} Zerill F J 1970 \emph{Phys.Rev.Let.} \textbf{24} 737
\bibitem{15} Ferrari V and Mashhoon V 1984 \emph{Phys.Rev.D} \textbf{30} 295
\bibitem{16} Nollert H P 1993 \emph{Phys.Rev.D} \textbf{47} 5253
\bibitem{17} Cardoso V and Lemos P S J 2001 \emph{Phys.Rev.D}
\textbf{64} 084017\\Zhidenko A 2004 \emph{Class.Quantum.Grav.}
\textbf{21} 273\\Cardoso V and Lemos P S J 2003 \emph{Phys.Rev.D}
\textbf{67} 084020
\bibitem{18} Leaver E W 1985 \emph{Proc.R.Soc.A} \textbf{402} 285
\bibitem{19} Konoplya R A 2002 \emph{Phys.Rev.D} \textbf{66} 044009
\bibitem{20} Khanal U 1985 \emph{Phys.Rev.D} \textbf{32} 879\\Park
M I 1998 \emph{Phys.Letters.B} \textbf{440} 275
\bibitem{21} Shao C C and Wang B \emph{et al} \emph{Phys.Rev.D} \textbf{71}
044003
\bibitem{22} Cardoso V and Lemos P S J 2001 \emph{Phys.Rev.D} \textbf{63} 124015\\Gupta K
S and Sen S 2005 \emph{Phys.Letters.B} \textbf{618} 237
\bibitem{23} Miranda A S and Zanchin V T 2006 \emph{Phys.Rev.D} \textbf{73}
064034\\Du D P \emph{et al} 2004 \emph{Phys.Rev.D} \textbf{70}
064024\\Govindarajan T R and Suneeta V 2001
\emph{Class.Quantum.Grav.} \textbf{18} 265\\Horowitz G T and
Hubeny V E 2000 \emph{Phys.Rev.D} \textbf{62} 024027
\bibitem{24}Chang J F and Shen Y G 2005 \emph{Nuclear.Phys.B}
\textbf{712} 347
\bibitem{25}Konoplya R A 2002 \emph{Phys.Rev.D} \textbf{66}
084007\\ Leaver E W 1990 \emph{Phys.Rev.D} \textbf{41} 2986\\
Konoplya R A 2002 \emph{Phys.Letters.B} \textbf{550} 117
\bibitem{26} Berti E and Kokkotas K D 2003 \emph{Phys.Rev.D}
\textbf{68} 044027\\ Hod S 2003 \emph{Phys.Rev.D} \textbf{67}
081501(R)
\bibitem{27} Jing J L and Pan Q Y 2005 \emph{Nuclear.Phys.B}
\textbf{728} 109\\ Berti E and Kokkotas K D 2005 gr-qc/0502065v2
\bibitem{28} Cho H T 2003 \emph{Phys.Rev.D} \textbf{68} 024003\\ Konoplya R A and Abdalla E 2005 \emph{Phys.Rev.D} \textbf{71}
084015\\ Nollert H P 1992 \emph{Phys.Rev.D} \textbf{47} 5253
\bibitem{29} Berti E \emph{et al} 2004 \emph{Phys.Rev.D}
\textbf{70} 124006\\ Chen S B and Jing J L 2005
\emph{Class.Quantum.Grav.} \textbf{22} 4651
\bibitem{30} Jing J L 2004 \emph{Phys.Rev.D} \textbf{69} 084009
\bibitem{31} Giammatteo M and Moss I G 2005 \emph{Class.Quantum.Grav.}
\textbf{22} 1803
\bibitem{32} Daghigh R G and Kunstatter G 2005 \emph{Class.Quantum.Grav.}
\textbf{22} 4113
\bibitem{33} Fiziev P P 2006 \emph{Class.Quantum.Grav.}
\textbf{23} 2447\\Konoplya R A 2005 \emph{Phys.Rev.D} \textbf{71}
024038\\ Ghosh A \emph{et al} 2006 \emph{Class.Quantum.Grav.}
\textbf{23} 1851
\bibitem{34} Konoplya R A 2002 \emph{Phys.Letters.B} \textbf{550}
117\\ Xue H L \emph{et al} 2002 \emph{Phys.Rev.D} \textbf{66}
024032
\bibitem{35} Simone L E and Will C M 1992 \emph{Class.Quantum.Grav.}
\textbf{9} 963
\bibitem{36} Burko L M and Khanna G 2004 \emph{Phys.Rev.D} \textbf{70}
044018
\bibitem{37} Konoplya R A and Zhidenko A V 2006 \emph{Phys.Rev.D}
\textbf{73} 124040


\bibitem{38}Kiselev V V 2003 \emph{Class.Quantum.Grav.}
             \textbf{20} 1187

\bibitem{39}Chen S B and Jing J L 2005 \emph{Class.Quantum.Grav.}
             \textbf{22} 4651







\bibitem{40}Schutz B F and Will C M 1985
\emph{Astrophy.J.Lett.Ed.} \textbf{291} L33
\bibitem{41}Iyer S and Will C M 1987 \emph{Phys.Rev.D}
            \textbf{35} 3621
\bibitem{42}Iyer S 1987 \emph{Phys.Rev.D}
            \textbf{35} 3632

\bibitem{43}Simone L E and Will C M 1992 \emph{Class.Quantum.Grav.}
             \textbf{9} 963


\end{thebibliography}
\end{document}